\documentclass[twocolumn,showpacs,preprintnumbers,amsmath,amssymb,superscriptaddress,pra]{revtex4-1}

\usepackage{graphicx}
\usepackage{dcolumn}
\usepackage{bm}
\usepackage{float}

\begin{document}


\title{High spatial entanglement via chirped quasi-phase-matched optical parametric
down-conversion}

\author{Ji\v{r}\'{i} Svozil\'{i}k}
\email{jiri.svozilik@icfo.es} \affiliation{ICFO---Institut de
Ciencies Fotoniques, Mediterranean Technology Park, 08860,
Castelldefels, Barcelona, Spain} \affiliation{Palack\'{y}
University, RCPTM, Joint Laboratory of Optics, 17.listopadu 12,
771 46 Olomouc, Czech Republic}

\author{Jan Pe\v{r}ina Jr.}
\affiliation{Palack\'{y} University, RCPTM, Joint Laboratory of
Optics, 17.listopadu 12, 771 46 Olomouc, Czech Republic}

\author{Juan P. Torres}
 \affiliation{ICFO---Institut de
Ciencies Fotoniques, Mediterranean Technology Park, 08860,
Castelldefels, Barcelona, Spain} \affiliation{Department of Signal
Theory and Communications, Universitat Politecnica Catalunya,
Campus Nord D3, 08034 Barcelona, Spain}

\date{\today}

\begin{abstract}
By making use of the spatial shape of paired photons, parametric
down-conversion allows the generation of two-photon entanglement
in a multidimensional Hilbert space. How much entanglement can be
generated in this way? In principle, the infinite-dimensional
nature of the spatial degree of freedom renders unbounded the
amount of entanglement available. However, in practice, the
specific configuration used, namely its geometry, the length of
the nonlinear crystal and the size of the pump beam, can severely
limit the value that could be achieved. Here we show that the use
of quasi-phase-matching engineering allows to increase the amount
of entanglement generated, reaching values of tens of ebits of
entropy of entanglement under different conditions. Our work thus
opens a way to fulfill the promise of generating massive spatial
entanglement under a diverse variety of circumstances, some more
favorable for its experimental implementation.
\end{abstract}

\maketitle

\section{Introduction}

Entanglement is a genuine quantum correlation between two or more
parties, with no analogue in classical physics.  During last
decades it has been recognized as a fundamental tool in
 several quantum information protocols, such as quantum teleportation \cite{bennett1993},
quantum cryptography \cite{ekert1991} and quantum key distribution
\cite{ribordy2000}, and distributed quantum computing
\cite{serafini2006}.

Nowadays, spontaneous parametric down-conversion (SPDC), a process
where the interaction of a strong pump beam with a nonlinear
crystal mediates the emission of two lower-frequency photons
(signal and idler), is a very convenient way to generate photonic
entanglement \cite{torres2011}. Photons generated in SPDC can
exhibit entanglement in the polarization degree of freedom
\cite{kwiat1995}, frequency \cite{law2000} and spatial shape
\cite{barbosa2000,mair2000}. One can also make use of a
combination of several degrees of freedom
\cite{barreiro2005,nagali2009}.

Two-photon entanglement in the polarization degree of freedom is
undoubtedly the most common type of generated entanglement, due
both to its simplicity, and that it suffices to demonstrate a
myriad of important quantum information applications. But the
amount of entanglement is restricted to $1$ ebit of entropy of
entanglement \cite{comment1}, since each photon of the pair can be
generally described by the superposition of two orthogonal
polarizations (two-dimensional Hilbert space).  On the other hand,
frequency and spatial entanglement occurs in an infinite
dimensional Hilbert space, offering thus the possibility to
implement entanglement that inherently lives in a higher
dimensional Hilbert space (qudits).

Entangling systems in higher dimensional systems (frequency and
spatial degrees of freedom) is important both for fundamental and
applied reasons. For example, noise and decoherence tend to
degrade quickly quantum correlations. However, theoretical
investigations predict that physical systems with increasing
dimensions can maintain non-classical correlations in the presence
of more hostile noise \cite{kaszlikowski2000,collins2002}. Higher
dimensional states can also exhibit unique outstanding features.
The potential of higher-dimensional quantum systems for practical
applications is clearly illustrated in the demonstration of the
so-called {\it quantum coin tossing}, where the power of higher
dimensional spaces is clearly visible \cite{molina2005}.

The amount of spatial entanglement generated depends of the SPDC
geometry used (collinear vs non-collinear), the length of the
nonlinear crystal ($L$) and the size of the pump beam ($w_0$). To
obtain an initial estimate, let us consider a collinear SPDC
geometry. Under certain approximations \cite{aprox}, the entropy
of entanglement can be calculated analytically. Its value can be
shown to depend on the ratio $L/L_d$, where $L_{d}=k_p w_0^2/2$ is
the Rayleigh range of the pump beam and $k_p$ is its longitudinal
wavenumber. Therefore, large values of the pump beam waist $w_0$
and short crystals are ingredients for generating high
entanglement \cite{oemrawsingh2005}. However, the use of shorter
crystals also reduce the total flux-rate of generated entangled
photon pairs. Moreover, certain applications might benefit from
the use of focused pump beams. For instance, for a $L=1$ mm long
stoichiometric lithium tantalate (SLT) crystal, with pump beam
waist $w_0=100$ $\mu$m, pump wavelength $\lambda_p=400$ nm and
extraordinary refractive index $n_e(\lambda_p)=2.27857$
\cite{bruner2003}, one obtains $E \sim 8.5$ \cite{aprox}. For a
longer crystal of $L=20$ mm, the amount of entanglement is
severely reduced to $E \sim 4.2$ ebits.

We put forward here a scheme to generate massive spatial
entanglement, i. e., an staggering large value of the entropy of
entanglement, independently of some relevant experimental
parameters such as the crystal length or the pump beam waist. This
would allow to reach even larger amounts of entanglement that
possible nowadays with the usual configurations used, or to attain
the same amount of entanglement but with other values of the
nonlinear crystal length or the pump beam waist better suited for
specific experiments.

\begin{figure}[t]
\includegraphics[width=6.5cm]{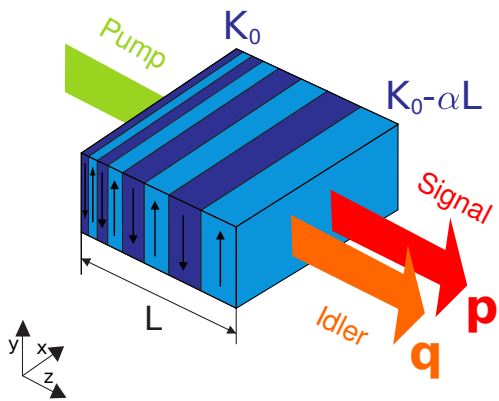}
\caption{Scheme of SPDC in a linearly chirped quasi-phase-matched
nonlinear crystal. The pump beam is a Gaussian beam, and ${\bf p}$
and ${\bf q}$ designate the transverse wave numbers of the signal
and idler photons, respectively. $K_0$ is the grating wave-vector
at the input face of the nonlinear crystal, and $K_0-\alpha L$ at
its output face. The signal and idler photons can have different
polarizations or frequencies.  The different colors (or different
direction of arrows) represent domains with different sign of the
nonlinear coefficient.} \label{Fig1}
\end{figure}

Our approach is based on a scheme originally used to increase the
bandwidth of parametric down-conversion
\cite{carrasco2004,nasr2008,svozilik2009}. A schematic view of the
SPDC configuration is shown in Fig.\ref{Fig1}. It makes use of
chirped quasi-phase-matching (QPM) gratings with a linearly
varying spatial frequency given by $K_{g}(z)=K_0-\alpha (z+L/2)$,
where $K_0$ is the grating's spatial frequency at its entrance
face ($z=-L/2$), and $\alpha$ is a parameter that represents the
degree of linear chirp. The period of the grating at distance $z$
is $p(z)=2\pi/K_g(z)$, so that the parameter $\alpha$ writes
\begin{equation}
\alpha=\frac{2\pi}{L} \frac{p_f-p_i}{p_f p_i}
\end{equation}
where $p_i$ is the period at the entrance face of the crystal, and
$p_f$ at its output face.

The key idea is that at different points along the nonlinear
crystal, signal and idler photons with different frequencies and
transverse wavenumbers can be generated, since the continuous
change of the period of the QPM gratings allows the fulfillment of
the phase-matching conditions for different frequencies and
transverse wavenumbers. If appropriately designed narrow-band
interference filters allow to neglect the frequency degree of
freedom of the two-photon state, the linearly chirped QPM grating
enhance only the number of spatial modes generated, leading to a
corresponding enhancement of the amount of generated spatial
entanglement.

\begin{figure}[t!]
\centering
\includegraphics[width=7.0cm]{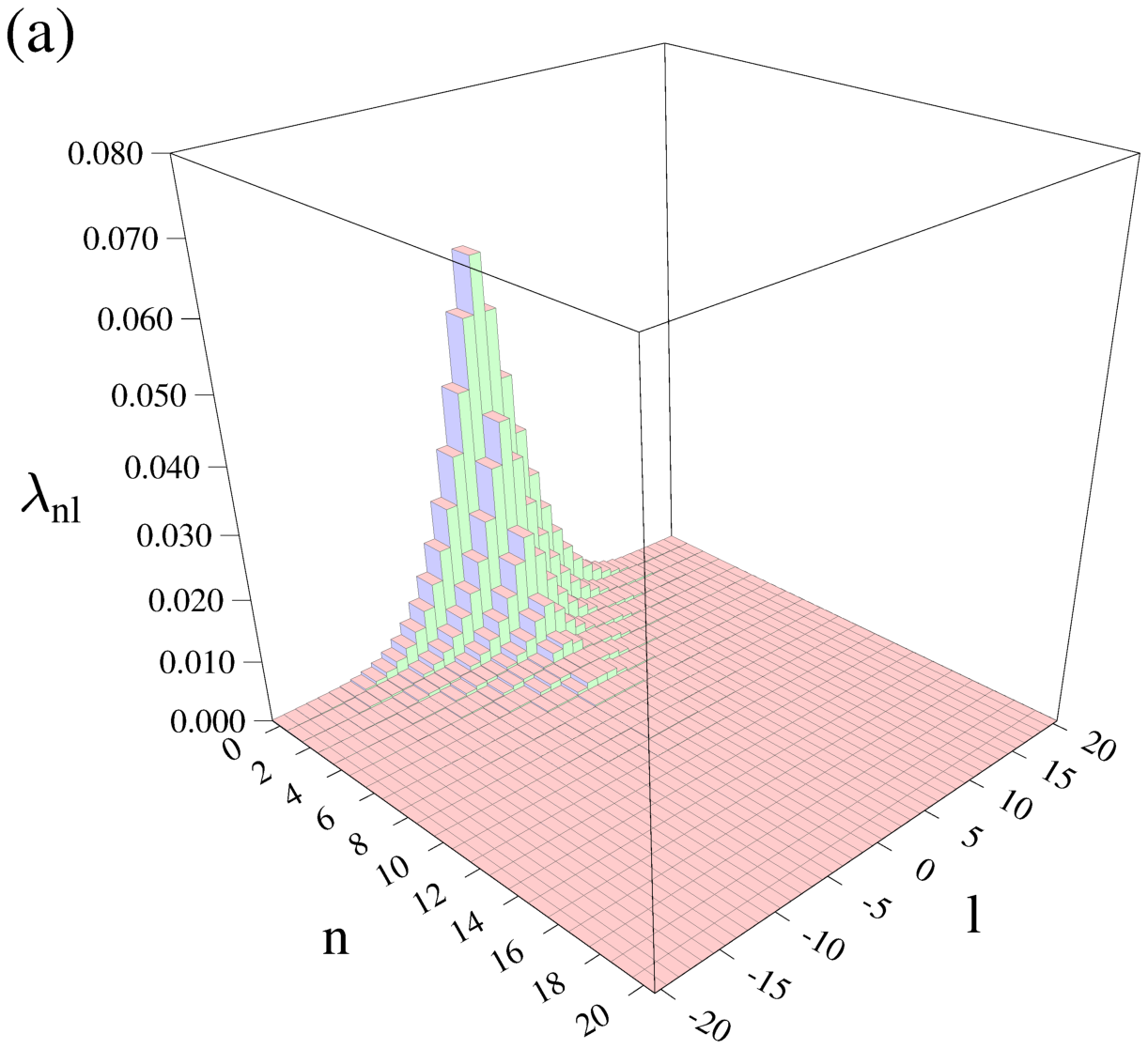}\\
\vspace{0.1cm}
\includegraphics[width=7.0cm]{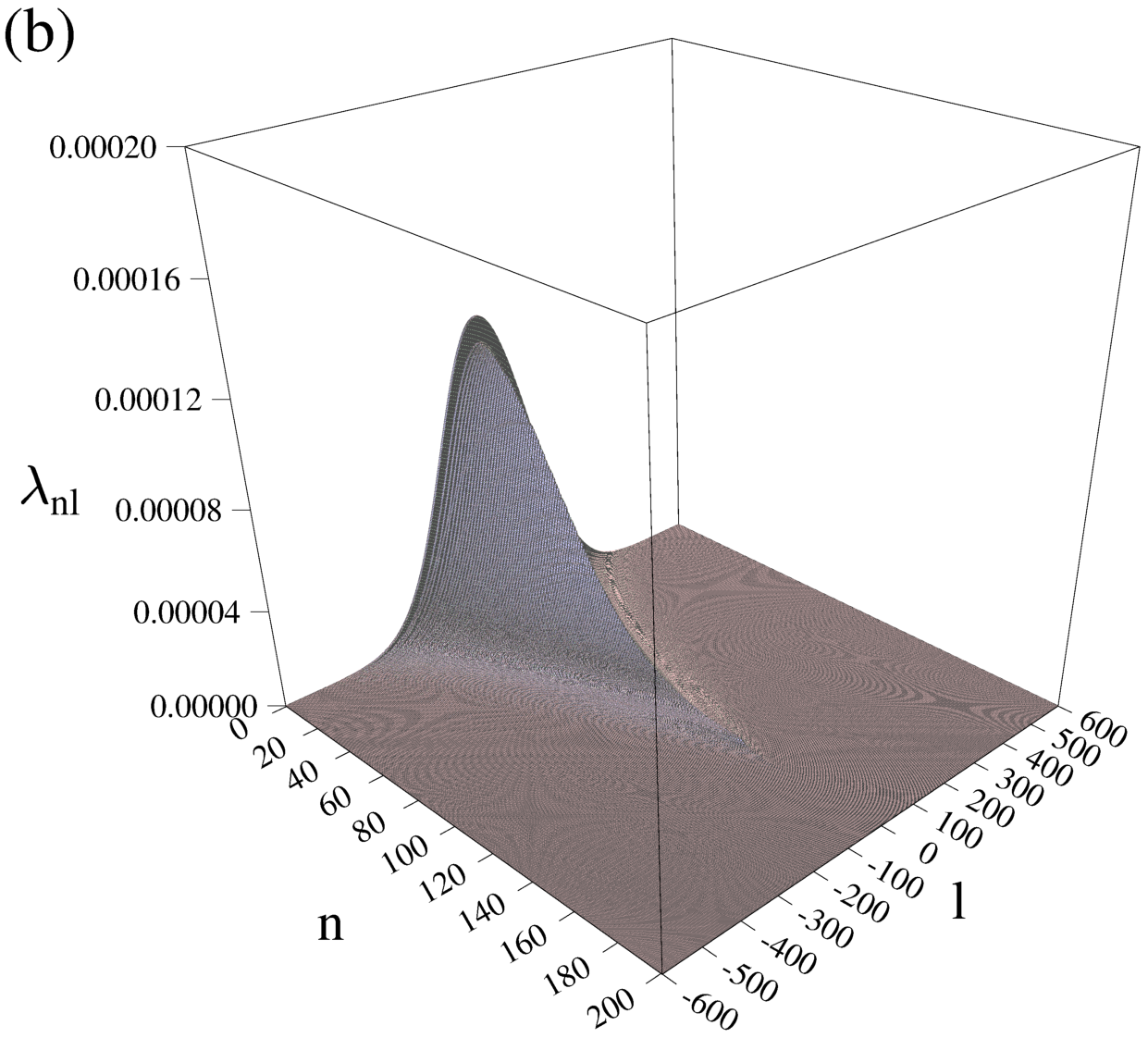}

\caption{Weight of the Schmidt coefficients $\lambda_{nl}$ for (a)
$\alpha=0\;\mu m^{-2}$ and (b) $\alpha=10\times 10^{-6}\;\mu
m^{-2}$. The nonlinear crystal length is $L=20$ mm and the pump
beam waist is $w_{0}=100\;\mu m$. } \label{Fig2}

\end{figure}

\section{Theoretical model}
In order to determine how much spatial entanglement can be
generated in SPDC with the use of chirped QPM, let us consider a
nonlinear optical crystal illuminated by a quasi-monochromatic
laser Gaussian pump beam of waist $w_0$. Under conditions of
collinear propagation of the pump, signal and idler photons with
no Poynting vector walk-off, which would be the case of a
noncritical type-II quasi-phase matched configuration, the
amplitude of the quantum state of the generated two-photon pair
generated in SPDC reads in transverse wavenumber space
\begin{equation}
|\Psi\rangle= \int d{\bf p} d{\bf q} \Psi \left({\bf p},{\bf
q}\right) |{\bf p} \rangle_s |{\bf q}\rangle_i, \label{state1}
\end{equation}
where ${\bf p}$ ($\bf q$) is the transverse wavenumber of the
signal (idler) photon. $\Psi$ is the joint amplitude of the
two-photon state, so that $|\Psi \left({\bf p},{\bf q}\right)|^2$
is the probability to detect a signal photon with transverse
wave-number ${\bf p}$ in coincidence with an idler photons with
${\bf q}$.

The joint amplitude that describes the quantum state of the paired
photons generated in a linearly chirped QPM crystal, using the
paraxial approximation, is equal to
\begin{eqnarray}
\label{state2}
\Psi \left( {\bf p},{\bf q}\right)&=&C \exp\left(-\frac{w^{2}_{0}}{4}|\mathbf{p+q}|^2\right)\\
&\times &\int_{-L/2}^{L/2}dz \exp \left[ i
\frac{|\mathbf{p}-\mathbf{q}|^2}{2 k_p}z+i \alpha \left(
z+\frac{L}{2} \right)z \right],\nonumber
\end{eqnarray}
where $C$ is a normalization constant ensuring $\int
d\mathbf{q}\int d\mathbf{p}| \Psi \left(
\mathbf{p},\mathbf{q}\right)|^2=1$. Notice that the value
of $K_0=2\pi/p(-L/2)$ does now show up in Eq. (\ref{state2}),
since we make use of the fact that there is phase matching for
${\bf p}={\bf q}=0$ at certain location inside the nonlinear
crystal, which in our case it turns out to be the input face
($z=-L/2$).

 After integration along the
z-axis one obtains

\begin{widetext}
\begin{eqnarray}
 \label{state3}
\Psi(\mathbf{p},\mathbf{q}) & = & C \sqrt{\frac{i\pi}{4\alpha}}exp\left[-\frac{w_{0}^{2}}{4}|\mathbf{p}+\mathbf{q}|^{2}-i\left(\frac{\alpha L^{2}}{16}+\frac{L|\mathbf{p}-\mathbf{q}|^{2}}{8k_{p}}+\frac{\mathbf{|p-}\mathbf{q}|^{4}}{16\alpha k_{p}^{2}}\right)\right]\nonumber \\
 &  & \times\left[ \mathrm{erf}\left(\frac{3\sqrt{\alpha}L}{4\sqrt{i}}+\frac{|\mathbf{p}-\mathbf{q}|^{2}}{4k_{p}\sqrt{i\alpha}}\right)-\mathrm{erf}\left(-\frac{\sqrt{\alpha}L}{4\sqrt{i}}+\frac{|\mathbf{p}-\mathbf{q}|^{2}}{4k_{p}\sqrt{i\alpha}}\right)\right],
\end{eqnarray}
\end{widetext}
where $\mathrm{erf}$ refers to the error function. Notice that Eq.
(\ref{state3}) is similar to the one describing the joint spectrum
of photon pairs in the frequency domain, when the spatial degree
of freedom is omitted \cite{nasr2008,svozilik2009}.
The reason is that both equations originate in phase matching
conditions along the propagation direction ($ z $ axis).

Since all the configuration parameters that define the down
conversion process show rotational symmetry along the propagation
direction $z$, the joint amplitude given by Eq. (\ref{state3}) can
be written as
\begin{equation}
\Psi\left(\mathbf{p},\mathbf{q}\right)=\sum^{\infty}_{l=-\infty}B_{l}\left(p,q\right)e^{i
l \left(\varphi_p-\varphi_q\right)}. \label{mode_decomposition}
\end{equation}
Here, we have made use of polar coordinates in the transverse
wave-vector domain for the signal, $\mathbf{p}=\left(p
\cos\varphi_p, p \sin\varphi_p\right)$, and idler photons
$\mathbf{q}=\left(q \cos\varphi_q,q \sin\varphi_q\right)$, where
$\varphi_{p,q}$ are the corresponding azimuthal angles, and $p,q$
are the radial coordinates. The specific dependence of the Schmidt
decomposition on the azimuthal variables $\varphi_p$ and
$\varphi_p$ reflects the conservation of orbital angular momentum
in this SPDC configuration \cite{osorio2008}, so that a signal
photon with OAM winding number $+l$ is always accompanied by a
corresponding idler photon with OAM winding number $-l$. The
probability of such coincidence detection for each value of $l$ is
the spiral spectrum \cite{torres2003} of the two-photon state,
i.e., the set of values $P_l=\int p dp\, q dq \,|B_l(p,q)|^2$.
Recently, the spiral spectrum of some selected SPDC configuration
have been measured \cite{pires2010}.

The Schmidt decomposition \cite{ekert1995,law2005} of the spiral
function, i.e.,
$B_l(p,q)=\sum_{n=0}^{\infty}\sqrt{\lambda_{nl}}f_{nl}(p)g_{nl}(q)$,
is the tool to quantify the amount of entanglement present.
$\lambda_{nl}$ are the Schmidt coefficients (eigenvalues), and the
modes $f_{nl}$ and $g_{nl}$ are the Schmidt modes (eigenvectors).
Here we obtain the Schmidt decomposition  by means of a
singular-value decomposition method. Once the Schmidt coefficients
are obtained, one can obtain the entropy of entanglement as
$E=-\sum_{nl}\lambda_{nl}\log_{2}\lambda_{nl}$. An estimation of
the overall number of spatial modes generated is obtained via the
Schmidt number $K=1/\sum_{nl}\lambda_{nl}^{2}$, which can be
interpreted as a measure of the effective dimensionality of the
system. Finally, the spiral spectrum is obtained as $P_l=\sum_n
\lambda_{nl}$.

\section{Discussion}

\begin{figure}[t]

\includegraphics[width=6.5cm]{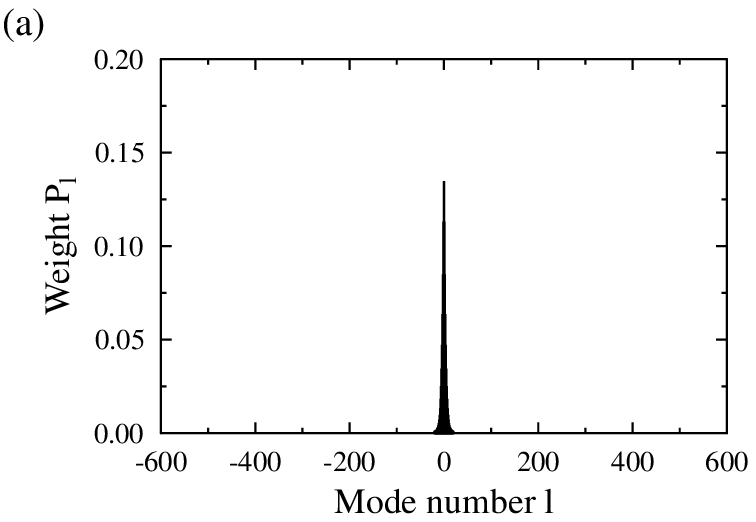} \\
\includegraphics[width=6.5cm]{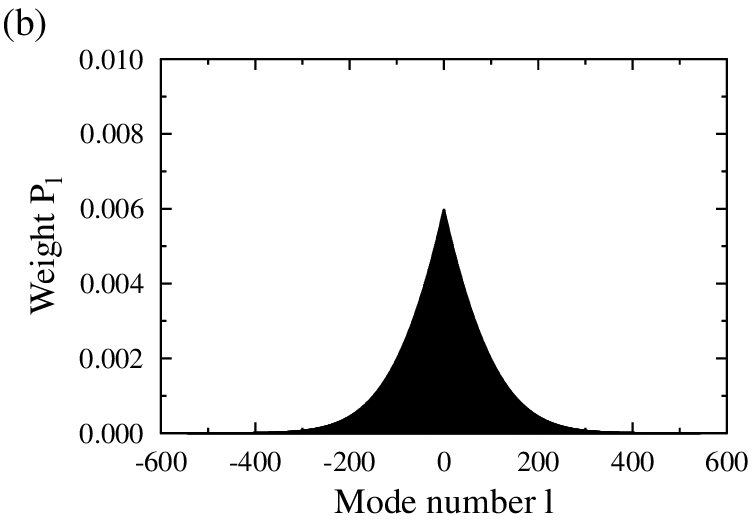}

\caption{The spiral spectrum $P_l$ for (a) $\alpha=0\;\mu m^{-2}$
and (b) $\alpha=10\times 10^{-6}\;\mu m^{-2}$. The nonlinear
crystal length is $L=20$ mm and the pump beam waist is
$w_{0}=100\;\mu m$. } \label{Fig3}
\end{figure}

For the sake of comparison, let us consider first the usual case
of a QPM SLT crystal with no chirp, i.e., $\alpha=\,0\,\mu m^{-2}$,
and length $L=20$ mm, pumped by a Gaussian beam with beam waist
$w_0=\,100\,\mu m$ and wavelength $\lambda_p=400$ nm. In this case, the integration of Eq.
(\ref{state2}) leads to \cite{walborn2003}
\begin{equation}
\Psi(\mathbf{p},\mathbf{q})=C
\exp{\left(-\frac{w^{2}_0}{4}|\mathbf{p}+\mathbf{q}|^{2}\right)}\mathrm{sinc}\left(\frac{L|\mathbf{p}-\mathbf{q}|^2}{4
k_p}\right). \label{state4}
\end{equation}
The Schmidt coefficients are plotted in Fig. \ref{Fig2}(a), and
the corresponding spiral spectrum is shown in Fig. \ref{Fig3}(a).
The main contribution to the spiral spectrum comes from the
spatial modes with $l=0$. The entropy of entanglement for this
case is $E = 6.4$ ebits and the Schmidt number is $K = 42.9$.

\begin{figure}[t]
\centering
\includegraphics[width=6.5cm]{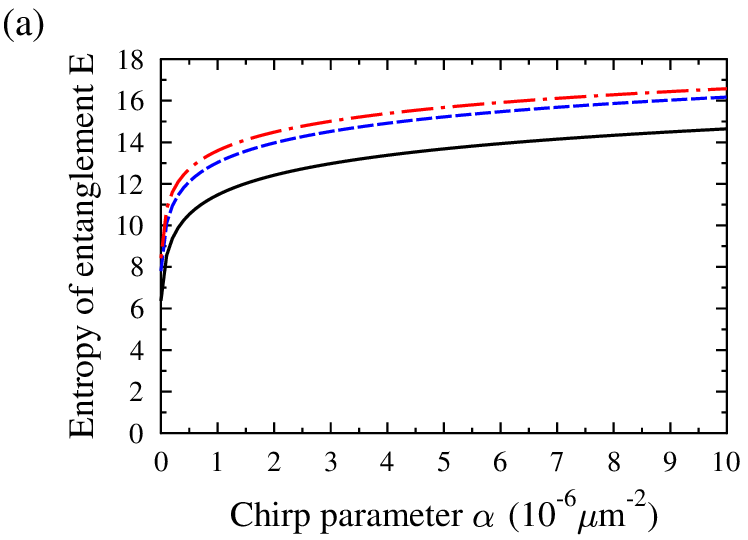}\\
\includegraphics[width=6.5cm]{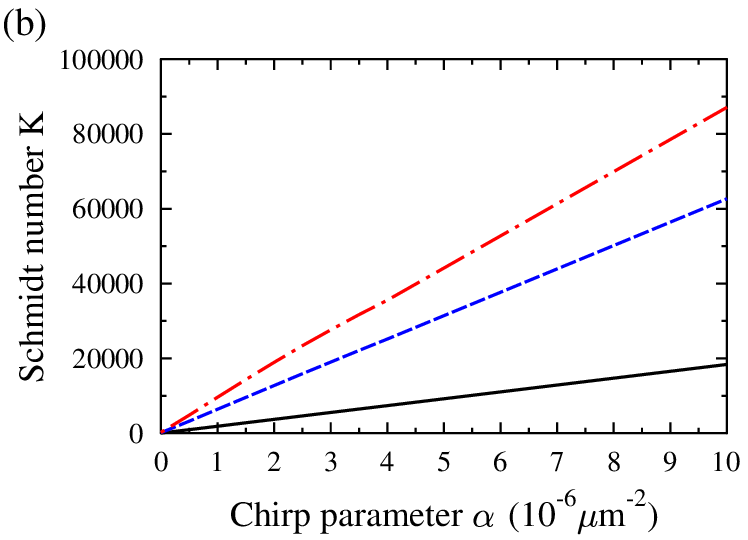}
\caption{(a) The entropy of entanglement $E$ and (b) the Schmidt
number $K$ as a function of the chirp parameter $\alpha$ for
$w_0=100\;\mu m$ (solid black line), $w_0=200\;\mu m$ (dashed blue
line) and $w_0=300\;\mu m$ (dotted-and-dashed red line). }
\label{Fig4}
\end{figure}

Nonzero values of the chirp parameter $\alpha$ lead to an increase
of number of generated modes, as it can be readily seen in Fig.
\ref{Fig2}(b) for $\alpha\,=10\times10^{-6}$ $\mu m^{-2}$ and
$w_0\,=100$ $\mu$m. This broadening effect is also reflected in
the corresponding broadening of spiral spectrum, as shown in Fig.
\ref{Fig3}(b). Indeed, Fig. \ref{Fig4}(a) shows that the entropy
of entanglement increases with increasingly larger values of the
chirping parameter, even though for a given value of $w_0$, its
increase saturates for large values of $\alpha$.  For
$w_0=\,300\,\mu m$ and $\alpha\,=10 \times  10^{-6}\,\mu m^{-2}$,
we reach a value of $E=16.6$ ebits. On the contrary, the Schmidt
number $K$ rises linearly with $\alpha$, as can be observed in
Fig. \ref{Fig4}(b), for all values of $w_0$. For sufficiently
large values of $w_0$ and $\alpha$, $K$ reaches values of several
thousands of spatial modes, i.e. $K=87113 $ for the same $w_0$ and
$\alpha$. For large values of $E$, a further increase of $E$
requires an even much larger increase of the number of spatial
modes involved, which explain why an increase of the number of
modes involved only produces a modest increase of the entropy of
entanglement. Notice that the spiral spectrum presented in Fig.
\ref{Fig3}(b) is discrete. Notwithstanding, it might look
continuous since it is the result from the presence of several
hundreds of OAM modes with slightly decreasing weights.

We have discussed entanglement in terms of transverse modes which
arise from the Schmidt decomposition of the two-photon amplitude
and, as such, they attain appreciable values in the whole
transverse plane. Alternatively, the existing spatial correlations
between the signal and idler photons can also be discussed using
second-order intensity correlation functions \cite{Hamar2010}. In
this approach, correlations are quantified by the size of the
correlated area ($\Delta {\bf p}$) where it is highly probable to
detect a signal photon provided that its idler twin has been
detected with a fixed transverse wave vector ${\bf q}$. We note
that the azimuthal width of correlated area decreases with the
increasing width of the distribution of Schmidt eigenvalues along
the OAM winding number $l$. On the other hand, the increasing
width of the distribution of Schmidt eigenvalues along the
remaining number $n$ results in a narrower radial extension of the
correlated area. An increase in the number of modes $K$ results in
a diminishing correlation area, both in the radial and azimuthal
directions. The correlated area drops to zero in the limit of
plane-wave pumping, where attains the form of a $\delta$ function.
The use of such correlations in parallel processing of information
represents the easiest way for the exploitation of massively
multi-mode character of the generated beams.

For the sake of comparison, when considering frequency
entanglement, the entropy of entanglement depends on the ratio
between the bandwidth of the pump beam (typically $B_p \sim 5$
MHz) and the bandwidth of the down-converted two-photon state
($B_{dc}$) \cite{parker2000,PerinaJr2008}. For type II SPDC, one
has typically values of $E \sim 1-2$ \cite{law2000}. Increasing
the bandwidth of the two-photon state, one can reach values of
$B_{dc}
> 1000$ THz, therefore allowing typical ratios greater than
$B_{dc}/B_p \gg 10^8$, with $E > 25$ \cite{martin2009}.

\section{Conclusion}

In conclusion, we have presented a new way to increase
significantly the amount of two-photon spatial entanglement
generated in SPDC by means of the use of chirped
quasi-phase-matching nonlinear crystals. This opens the door to
the generation of high entanglement under various experimental
conditions, such as different crystal lengths and sizes of the
pump beam.

QPM engineering can also be an enabling tool to generate truly
massive spatial entanglement, with state of the art QPM
technologies \cite{nasr2008} potentially allowing to reach
entropies of entanglement of tens of ebits. Therefore, the promise
of reaching extremely high degrees of entanglement, offered by the
use of the spatial degree of freedom, can be fulfilled with the
scheme put forward here. The experimental tools required are
available nowadays. The use of extremely high degrees of spatial
entanglement, as consider here, would demand the implementation of
high aperture optical systems. For instance, for a spatial
bandwidth of $\Delta {\bf p} \sim 2 \mu m^{-1}$, the aperture
required for $\lambda_p=400$ nm is $\Delta \theta \sim
4^{\circ}-6^{\circ}$.

The shaping of QPM gratings are commonly used in the area of
non-linear optics for multiple applications such as beam and pulse
shaping, harmonic generation and all-optical processing
\cite{hum2007}. In the realm of quantum optics, its uses are not
so widespread, even though QPM engineering has been considered,
and experimentally demonstrated, as a tool for spatial
\cite{qpm2004,yu2008} and frequency \cite{nasr2008} control of
entangled photons. In view of the results obtained here concerning
the enhancement of the degree of spatial entanglement, it could be
possible to devise new types of gratings that turn out to be
beneficial for other applications in the area of quantum optics.

This work was supported by the Government of Spain (Project
FIS2010-14831) and the European union (Project PHORBITECH,
FET-Open 255914). J. S. thanks the project FI-DGR 2011 of the
Catalan Government. This work has also supported in part by
projects COST OC 09026, CZ.1.05/2.1.00/03.0058 of the Ministry of
Education, Youth and Sports of the Czech Republic and by project
PrF-2012-003 of Palack\'{y} University.


\begin{thebibliography}{99}
\bibitem{comment1} For a two-photon state with density matrix $\rho_{12}$, the entropy of entanglement is defined as
$E=-Tr \left( \rho_1 \log_2 \rho_1 \right)$, where $\rho_1= Tr_2
\,\rho_{12}$ is the partial trace over the variables describing
subsystem $2$ of the global density matrix. The entropy of
entanglement of a maximally entangled quantum state, whose two
parties live in a $d$-dimensional system, is $log_2 d$. Since the
state of polarization of a single photon is a two-dimensional
system, the maximum entropy of entanglement is $1$.
\end{thebibliography}
\end{document}